%% file: main.tex
\documentclass[10pt, conference, letterpaper]{IEEEtran}

\usepackage{epsfig,endnotes,multirow}
\usepackage{algorithmic,algorithm}

\usepackage{color}
\usepackage{bbm}
\usepackage{fancyvrb}
\usepackage{cleveref}
\usepackage{hyperref}
\usepackage[T1]{fontenc}
\usepackage{url}

\crefformat{footnote}{#2\footnotemark[#1]#3}

\definecolor{mygreen}{rgb}{0,0.5,0}
\definecolor{myred}{rgb}{0.5,0,0}
\definecolor{myblue}{rgb}{0,0,1}
\definecolor{myorange}{rgb}{1,0.65,0}

\newcommand{\anatv}[1]{{#1}}
\newcommand{\anat}[1]{{#1}}
\newcommand{\david}[1]{{#1}}

\newcommand{\iotica}{{\sc IoTica}~}

\newcommand\ignore[1]{}
\newcommand\ignorenoms[1]{}
\newcommand{\negs}{\hspace{-0.25em}}

\usepackage{epsfig,endnotes}

\usepackage{cite}
\usepackage{rotating}

\begin{document}
\thispagestyle{plain}


\title{NFV-based IoT Security for Home Networks using MUD}

\author{
Yehuda Afek\IEEEauthorrefmark{1}, Anat Bremler-Barr\IEEEauthorrefmark{3},  David Hay\IEEEauthorrefmark{2}, Ran Goldschmidt\IEEEauthorrefmark{3}, Lior Shafir\IEEEauthorrefmark{1},  Gafnit Abraham\IEEEauthorrefmark{3}, Avraham Shalev\IEEEauthorrefmark{3} \\ \ \\
   \IEEEauthorblockA{\IEEEauthorrefmark{1}The School of Computer Science, Tel Aviv University, Tel Aviv, Israel.}
   \IEEEauthorblockA{\IEEEauthorrefmark{2}School of Engineering and Computer Science, Hebrew University, Jerusalem, 
  Israel.}
   \IEEEauthorblockA{\IEEEauthorrefmark{3}Computer Science Department, Interdisciplinary Center, Herzliya, Israel.} 

}

\ignore{
\author{
{\rm  }\\
Tel-Aviv University
\and
{\rm Anat Bremler-Barr}\\
Interdisciplinary Center
\and
{\rm David Hay}\\
Hebrew University
\and
{\rm Ran Goldschmidt}\\
Interdisciplinary Center
\and
{\rm Lior Shafir }\\
Tel-Aviv University\\
\and
{\rm Gafnit Abraham }\\
 Interdisciplinary Center\\
\and
{\rm Avraham Shalev}\\
Interdisciplinary Center
}

}







\maketitle

\begin{abstract}
 \input{abstract}

\end{abstract}

\section{Introduction}
\input{IntroNew.tex}
\section{NFV-based System Description}

\input{SystemIoTSec.tex}

\input{P2P.tex}

\section{Implementation and PoC Details} 
\input{implement}
\section{Discussion and Future Work}

\input{Discussion}
{\footnotesize 
\bibliographystyle{IEEEtran}
\bibliography{sample}

}

\end{document}

%% file: abstract.tex
A new scalable ISP level system architecture to secure and protect all IoT devices in a large number of homes is presented.
The system is based on whitelisting, as in the Manufacturer Usage Description (MUD) framework, implemented as a VNF. Unlike common MUD suggestions that place the whitelist application at the home/enterprise network, our approach is to place the enforcement upstream at the provider network, combining an NFV (Network Function Virtualization)  with router/switching filtering capabilities, e.g., ACLs. The VNF monitors many home networks simultaneously, and therefore, is a highly-scalable managed service solution that provides both the end customers and the ISP with excellent visibility and security of the IoT devices at the customer premises.  

The system includes a mechanism to distinguish between flows of different devices at the ISP level despite the fact that most home networks (and their IoT devices) are behind a NAT and all the flows from the same home come out with the same source IP address.
Moreover, the NFV system needs to receive only the first packet of each connection at the VNF, and rules space is proportional to the number of unique types of IoT devices rather than the number of  IoT devices. The monitoring part of the solution is off the critical path and can also uniquely protect from incoming DDoS attacks.

To cope with internal traffic, that is not visible outside the customer premise and often consists of P2P communication, we suggest a hybrid approach, where we deploy a lightweight component at the CPE, whose sole purpose is to monitor P2P communication. As current MUD solution does not provide a secure solution to P2P communication, we also extend the MUD protocol to  deal also with peer-to-peer communicating devices.

A  PoC  with a  large  national  level  ISP  proves  that  our  technology  works  as expected, identifying the various IoT devices that are connected to the network and detecting any unauthorized communications.

%% file: IntroNew.tex
IoT devices are strong enough to host malicious code or become a zombie, but, for economic and technological reasons, they do not have the means to protect themselves from being hacked. Thus, the billions of IoT devices (estimates are from 10 to 50 billion by 2020, 50\% being consumer devices) are fertile ground for many attacks of different kinds: DDoS attacks, privacy infringing attacks, data leakage, and physical break-ins. Several different systems and methods to provide protection to IoT devices have been proposed in recent years \cite{5,6,7,Meidan:2017:PML:3019612.3019878,IoTcities,DDoS_IoT}. In this paper, we present an ISP level system that protects the IoT devices in a large number of homes. This effective, scalable system should be a key part of protecting and stopping attacks on and from IoT devices. 

The underlying property that distinguishes most IoT devices from home computers, such as laptops, desktops, and hand-held devices, is that in most cases an IoT device communicates with a limited number (most usually very small) of pre-defined domains. Furthermore, these devices, rarely if ever, go through a firmware update. These characteristics have been exploited to create a whitelist based protection methods for IoT devices. A recent initiative that receives a lot of attention, both in industry and academia, calls for IoT device vendors to provide a \emph{Manufacturer Usage Description (MUD)} for their products~\cite{ietf-opsawg-mud-25}. These descriptions, called MUD files, consist of whitelists describing the devices' legitimate communication, specified by the domain names of legitimate endpoints. Methods to automatically learn and acquire the MUD file for a device type have also been developed \cite{DBLP:journals/corr/abs-1804-04358}.

To protect an IoT device, given its MUD file, or any other whitelist, it is necessary to resolve the domain names in the list to get the corresponding IP addresses (which might vary in different geographies and at different times), then to monitor the device traffic ensuring it communicates only with these IP addresses and complies with the MUD file (e.g., uses the specific ports and protocols specified in the file). If a deviation is detected, offending connections should be blocked and alerts on suspicious activity are issued.  Suggested implementations of the MUD standard as well as other IoT protection systems, implement these steps within the internal/local area network (LAN), i.e., for home networks on the CPE, \cite{MCAFEE,ASUS,3,Heidmall,1,6,7,5,2,4}, though sometimes on a separate device. In this paper, we provide a system architecture to implement these steps at the ISP level. The system combines an off-path Virtual Network Function (VNF) with the existing ISP's on-path enforcement capabilities. The VNF holds the MUD rules (where domain names are already resolved to IP addresses) and monitors a large number of home networks by these rules. Upon a violation, the corresponding connection is blocked in one of the ISP's on-path routers or switches, e.g., using ACLs. 

As home networks apply Network Address Translation (NAT), implying all devices use the same IP address and port numbers are arbitrary, perhaps the major challenge when working outside the LAN is to distinguish between connections originating from different devices. Our design overcomes this difficulty by dynamically installing \emph{packet marking rules} on the home gateway router (often called a Customer Premise Equipment or CPE), using its standard configuration protocol TR-69\cite{tr69}. Marking is done on the DSCP header field (thus not introducing extra overhead) only on packets originating from IoT devices. Furthermore, marking rules are installed only when the device first connects to the CPE.

The IoT protection architecture suggested here has several advantages: First, it is not on the critical path of the traffic as the VNF takes only a copy of the first packet of each connection. Second, for any home whose CPE is managed by the ISP (the vast majority of homes) no cooperation from the client is necessary, the protection may be provided as a service in one click on the ISP management unit. 
Third, our system provides the customer and the ISP a complete
view of the devices in the network. This is achieved by retrieving information from the CPE using TR-069 and identification
of the IoT devices using existing MUD files (or other whitelists) installed for devices across the entire network.  Moreover, the
customer can receive an alert for any device that connects
to its network. These views of customer devices are also an
important and helpful service for customer help service at the
ISP.
Fourth, the number of rules held in the VNF is proportional to the number of IoT device types (unique MUD files), and not to the total number of IoT devices it is monitoring; moreover, domain resolution is done in the VNF once for all devices of the same type, thus reducing the load on the DNS server and enable using more secure DNS servers. 
The centralized, upstream location of the solution, observing many home networks at once, holds a great promise as the system can correlate traffic surges and patterns from multiple homes to detect and mitigate outgoing DDoS attacks and other global attacks. Furthermore, the system is situated at a good point to stop additional attack vectors, such as attacks on (or that come out of) the CPE itself\footnote{Our system also monitors all connections that terminate in the CPE, as the CPE itself is considered an IoT device} or incoming DDoS attacks on IoT devices and homes.


\ignorenoms{
We note that a limitation of our architecture is that it does not observe internal traffic within the home network, and might miss attacks on IoT devices from other internal devices in the same home. For such attack vectors, we suggest to install a lightweight component on the CPE itself that is in charge on monitoring \emph{only} intra-LAN traffic.}


To cope with internal traffic, that is not visible outside the customer premise, we suggest a hybrid approach, where we deploy a lightweight component at the CPE, whose sole purpose is to monitor such traffic. 
We extend the MUD specification and its whitelisting capabilities to support P2P communication in which IoT devices act as servers and their users (e.g., another device within the local network) act as clients. P2P communication is most common in intra-LAN communication. Yet, our system also covers users' remote access (e.g., using a mobile phone), which can be granted using port forwarding on the CPE, UPnP\cite{boucadair2013universal}, or STUN\cite{rosenberg2003stun} with Hole-Punching \cite{ford2005peer}. We note that 
extending the MUD protocol to support such communication, and then to incorporate its monitoring and enforcement in the ISP level, is of great importance since recently-reported botnets, like MIRAI~\cite{MIRAI}, targeted devices based on P2P communication.
 Notice that as reported in Shodan \cite{Shadon}, there are still many IoT devices that are vulnerable. Besides covering more attack vectors, such a hybrid system is beneficial to provide a certain level of security even if one of its components fails. 
 
 The rest of the paper is organized as follows. In Section~\ref{sec:related} we will discuss related work on MUD and IoT whitelisting. Section~\ref{sec:sys} describes our ISP-level system, which includes a virtual network function, whitelist enforcement component, and communication with the CPE to ensure correct packet marking. Section~\ref{sec:p2p} describes how we extend the MUD specification to support P2P communication (either internal traffic within the home network or P2P communication across the Internet). Section~\ref{sec:implement} provides details on our implementation and our PoC deployment with a large national-level ISP. Concluding remarks are given in Section~\ref{sec:discussion}.


\section{Related Work}
\label{sec:related}

 MUD (Manufacturer Usage Description) \cite{ietf-opsawg-mud-25}, has been officially approved to define IoT devices communication patterns  in order to reduce the attack surface on IoT devices.  In this standard, IoT vendors/providers are encouraged to provide a MUD file consisting of access control rules that describe the device's proper communication behavior \cite{jethanandani2018network}. Recently,  Cisco announced that some of devices support MUD and some vendors already defined MUD files \cite{MUDCisco}.  The MUD framework assume that the components at the network are MUD controller, that process the MUD information, and the router/switch that run ACL (using a whitelist approach). Thus, in the home-network environment, it assumes that the CPE deploys MUD.

In contrast, our system is deployed within the ISP network and receives a copy of the traffic from a PoP router. As far as we are aware, this is the only MUD solution that is deployed at the ISP level. Moreover, all IoT security frameworks,  including the ones that do not use MUD or whitelisting at all, that we are aware of (either from Academia or Industry) focus on solutions
within the CPE/LAN and not in the ISP level~\cite{5,6,7,Meidan:2017:PML:3019612.3019878,IoTcities,DDoS_IoT,1,3,Heidmall,8366982}. 

 MUD-based solutions are still incomplete, and there are several recent works  that focus on extending or implementing  MUD, some of them in SDN environment \cite{ranganathan2019soft,Hamza:2018:CMP:3229565.3229571,Hamza:2019:DVA:3314148.3314352}. Important issues that are yet to be addressed are dealing with IoT devices that may be accessed directly by the owner mobile device(s) (in a P2P manner), dealing with services behind a cloud bucket (e.g., Amazon S3 bucket), where the exact bucket is specified in an encrypted manner, and dealing with devices in which the user can add apps/skills (such as a smart TV or Amazon Echo). \anatv{In this paper we extend MUD to handle P2P traffic.}

%% file: SystemIoTSec.tex
\label{sec:sys}

The system goal is to ensure that all packets from and to an IoT device comply with the MUD file rules. This implies that, for each packet , the system needs to decide whether it conforms with a MUD file (or another form of a whitelist), and if not, to block the packet. The MUD enforcement has thus two logical components, monitoring and enforcing. In the \emph{whitelist monitoring} (WLM), it is determined whether a packet/connection complies with a whitelist/MUD file or not; In the \emph{whitelist/MUD enforcement} (WLE), based on the output of the WLM component, the packet is either dropped or permitted. While WLE must be \emph{on the traffic path} observing every packet originated or destined for IoT devices; the WLM may be performed on a copy of the traffic, thus being off the critical path. In our implementation WLM, is implemented as a VNF that receives only a single packet of each connection, sent from an IoT device, to decide whether the entire 
connection should be permitted or blocked.

\begin{figure}
\includegraphics[clip, trim=0cm 0.65cm 0cm 1.85cm, width=0.5\textwidth]{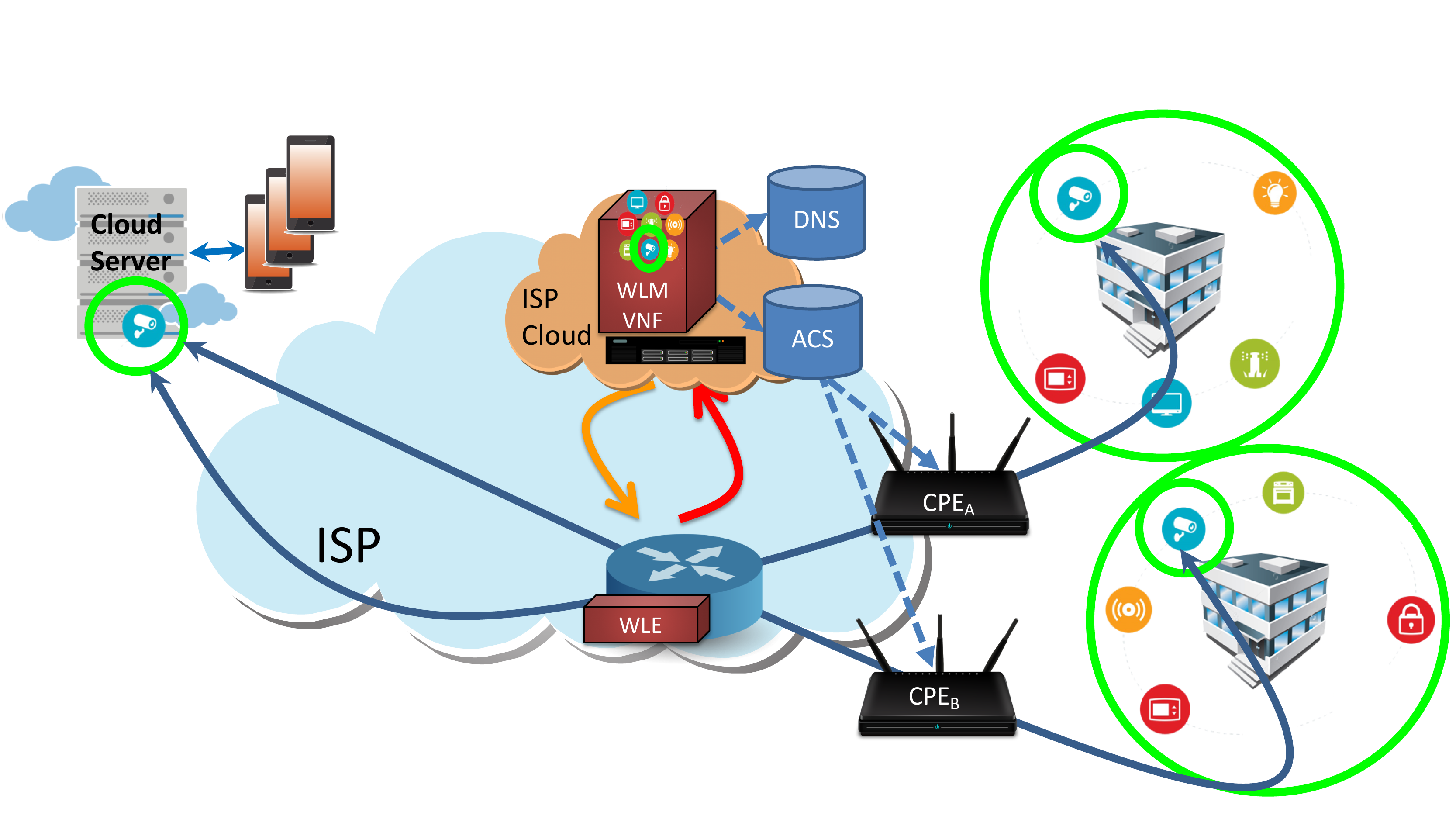}
\caption{\label{fig:framework}  Illustration of our Framework. The WLM's data plane is marked with solid lines, while the WLM's control plane is marked with dash lines. In this illustration, the WLM VNF protects two LANs that communicate with the Internet through two CPEs. Both LANs have a camera installed in them, which connects to the same service in the cloud. The WLM VNF contains whitelists/MUD files for all IoT devices in the system. In this particular case, it verifies that both cameras communicate only with endpoints within their whitelist/MUD file.}
\end{figure}

The control-plane in our implementation includes the control-plane inside the VNF as well as control communication with other control-planes, such as on the CPEs and other components, see Figure~\ref{fig:framework}.
We next describe its operation when the IoT devices act as clients  initiating connections to their cloud services (or other services). We assume that customers connect to the ISP network through a CPE with a unique IP address (which can change over time). As in the typical setting, the IoT devices reside behind a NAT (performed by the CPE), implying that outside the customer network all outgoing packets appear to come from the same IP address with arbitrary ports. 
\anat{We \david{further} assume that IoT devices have \david{a} MUD \david{file} (or \david{another readily-available} whitelist). \david{Notice that our framework also provides} the infrastructure to \david{\emph{learn}} a new MUD \david{file} of devices \david{whose} manufacturer did not \david{provide one}. The \david{method} itself of learning the MUD \david{files} from deployed IoT \david{devices} in the network is out of the scope of this paper. We note that there are tools like MUDgee \cite{DBLP:journals/corr/abs-1804-04358}  that create a MUD file from traffic \david{traces of} labeled devices in the lab. However, it is not straightforward to apply these tools on \david{ unlabeled IoT devices, that are deployed in the wild, and one has no control over their operations (specifically, one cannot actively try specific device functionality and observe the resulting traffic)}.  Nevertheless,  we note that the centralized view of our system is better in learning the regular behaviour of IoT devices (i.e., \david{generating a} MUD \david{file}), since it can observe and analyze multiple devices \david{of the same type}. }

\subsection{Data Plane}
\label{sec:dataplane}

The data-plane has three major parts: the customers' private networks, the ISP network, and the VNF, which monitors traffic and issues alerts (on suspicious traffic). We assume that a MUD file, containing a whitelist for each IoT device \emph{type} $M$, is given as a set of domains $WLD_M$, that correspond to IP addresses $WL_M$ with which the device is allowed to communicate; namely, $WL_M=\bigcup_{D\in WLD_M}\{L\in\mbox{DNS-response}(D)\}$. Moreover, $WL_M$ may contain more parameters (such as port numbers) as defined by the MUD protocol specification. We note, however, that our framework does not support Layer-2 parameters (such as MAC address) as these are not visible outside the LAN. 
\ignore{These parameters are rarely used anyway, because MUD files are common for all devices of a specific type, and therefore, should not contain specific physical addresses.  }

\subsubsection{Operations within the Customers' Network}

IoT devices that are monitored by the framework are oblivious to its existence, and no special configuration needs to be done at the IoT device level. 
However, to be able to distinguish packets coming from different IoT devices at the ISP level, CPEs that are connected to our framework should \emph{mark} packets originating from different IoT devices with different values in the DSCP field in the IP packet header. DSCP marking (by source MAC address) is a standard feature in most CPEs available today and requires only CPE \emph{configuration}. We note that packets originating from non-IoT devices (such as desktops, laptops, or mobile phones) are not marked with a special DSCP value (and maintain the original value used by the device; $0$ in most cases). A newly-connected device, for which there is no indication whether it is an IoT device or not, is marked and tracked until such a decision is made.

The remote CPE configuration (namely, the assignment of a DSCP value for each device) is part of the control plane's operations as described in Section~\ref{sec:controlplane}.


\subsubsection{Operations within the ISP network}
After transmitted from the customer premise, packets are sent through a sequence of routers within the ISP network, towards their  destination. Our framework requires that one router along this route will \emph{copy} outgoing packets from all monitored customer premises and send them to the VNF. This can be done either using port mirroring or installing a specific filter on the ISP's router. Moreover, the VNF needs only \emph{one packet from each outgoing connection}\footnote{We define a connection in this context as all packets that share the same $\langle src\_ip, src\_port, dst\_ip, dst\_port, protocol\rangle$ tuple.}.
In case the router supports the OpenFlow protocols, for example, this can be done by dynamically installing a filter on every observed connection. 

We note that packets are \emph{copied} (and not routed) from the router to the VNF, thus our VNF does not reside on the critical path of the packets  and does not induce extra delay or other performance penalties on the traffic.

The VNF may issue either alert on abnormal behavior or access-control list (ACL) filters to block malicious traffic in the WLE. The ACLs can be either specific  (e.g., block a specific connection; in most cases, both directions of the connection are blocked simultaneously) or aggregated  (block all traffic to/from specific address/network; this is especially appealing in case of DDoS attack where many malicious connections are created simultaneously).

It is important to notice that our WLM VNF observes only traffic originating from an IoT device (and therefore, is marked by the CPE). Since most known attack traffic is bidirectional \anatv{(e.g., for TCP traffic connections,  at least acknowledgments are sent)   }, our WLM VNF can detect it. As mentioned before, our WLE component blocks traffic from both directions. 

\subsubsection{Operations within the VNF}
\label{sec:dp_guard}
Packets received by the VNF are treated by their DSCP value, where classification actions are mapped to the multiple match-action table (mMT) model (e.g., as in OVS~\cite{ovs}). The exact mapping is described in Figure~\ref{table:tables}; table rules are updated by the control plane (see Section~\ref{sec:controlplane}). Specifically, we distinguish between the following cases:  

\smallskip
\noindent
{\bf Unmarked packets:}  Packets that have a DSCP value of $0$ (or another commonly-used value) were originated from a device that is not an IoT device, and therefore, is not monitored by our framework, implying our VNF should discard these packets.  

\noindent
{\bf Marked packets of an identified IoT device:} When a device joins a customer network, the CPE is configured to mark all its packet with a certain value. Through our control plane, we try to obtain the MUD file corresponding to this device and associate it with its specific type $X$. 
This implies that one should only verify that the packets comply with the corresponding MUD file/whitelist. 
This is done by matching  the header of the packet to $WL_X$ (e.g., verifying that the destination IP address is in the whitelist). In case any deviation is detected, an alert is issued and/or the connection is blocked in the WLE.


\noindent
{\bf Marked packets of an unidentified device:} 
In case no MUD file exists, we collect all of the device DNS queries and all the endpoints the device is communicating with.  This information is sent to the control plane that analyzes this information and decides which further actions should be taken. See Section~\ref{sec:controlplane_analyze} for more details. 

We note that due to technical reasons, there might be a gap between the time a device joins a customer's network and the time the corresponding CPE starts marking its packets. In order not to treat these first packets as unmarked, we configure the CPE to have a default proprietary DSCP value for all packets originating from the CPE. All such packets are treated as marked packets of an unidentified device, and the VNF associates them with the recently joined device (if only one such device exists).  See Section~\ref{sec:control_acs} for more details. 

It is important to notice that we store only one whitelist (or MUD file) per IoT device type (namely, all IP cameras of a specific model are monitored with a single whitelist in memory, regardless of their number). 
Specifically, if there are 21 commonly-used DSCP values~\cite{commondscp}, $C$ customers in the framework, each with $d_i$ IoT (or unidentified) devices, and $P$ whitelist/MUD-file, each with $p_j$ entries (e.g., legitimate destination IP address, port number, etc.), then the total number of filters required to implement the entire  data-plane is 
$
23+\sum_{i=1}^C (d_i{+}1) + \sum_{j=1}^P (p_j{+}1).
$
Furthermore, all update events (such as a change in the external IP address of a customer, new device has joined a customer's network, a profile has a new legitimate IP address, or a device is identified as a non-IoT device) requires a single filter update (in a single table within the pipeline).

\begin{figure}[tb]
\small{
\begin{tabular}{|l|l|l|}
\hline 
Table & Match & Action \\
\hline \hline 
0 & DSCP value is  & drop \\ 
& commonly-used & \\ \cline{2-3}
& otherwise & goto Table 1 \\
\hline
1 & source IP address is  & apply unique metadata,  \\
& of a customer &   goto Table 2 \\
\hline
2 & Meta-data, DSCP value & unidentified device: \\
& & send to controller \\ \cline{3-3}
&  & identified device (type \\
& & $X$): goto Table $X$ \\
\hline
$X{\geq}3$ & destination IP in $WL_X$ & drop \\
 \cline{2-3}
& (dest port, protocol) in $WL_X$ & drop \\
 \cline{2-3}
& otherwise & Output 1 \\
\hline
\end{tabular}
}
\caption{\label{table:tables} Our pipelined match-action implementation. An application that deals with traffic that violates the whitelist (e..g, generates ACL and sends it to the WLE) is connected to Output 1. In case the number of types exceeds the number of available tables, several types can reside on the same table, by applying a meta-data value to distinguish between them in Table 2.  Counters of filters in Table 2 present per-device activity of each customer. }
\end{figure}



\subsection{Control Plane}
\label{sec:controlplane}


The control plane operations are divided into the following tasks that are executed periodically, when a new customer joins the framework, or when a new device joins an existing customer's network (namely, a new device is connected to the network for the first time). 

\subsubsection{Maintaining data plane's whitelists}
\label{sec:activedns}
The first task of the control plane is to retrieve, for each IoT type $M$, the whitelist $WL_M$ from the given domain list $WLD_M$ that is specified within its MUD file.

One possibility is by looking at the DNS responses to the DNS queries it already tracks. However, this implies that both traffic directions should be monitored (sometimes this is not feasible) and the whitelist is prone to contamination, e.g., by DNS poisoning attacks, as the DNS response might be bogus.

We chose an alternative way of performing \emph{active DNS queries} of the domains in all the $WLD$ lists. This is done periodically (every few minutes) to keep with changes in IP addresses.  Moreover, upon the detection of communication of IoT device of type $M$ to an unidentified endpoint,  DNS queries of all  $WLD_M$ are immediately triggered. This is done to verify that IP addresses did not change since the last active DNS query and thus eliminating false positives alerts.  We take extra care  to avoid learning malicious IP addresses due to DNS poisoning, by performing the DNS queries to a secure DNS server (such as OpenDNS or Quad9) that is immune to such attacks. It is important to notice that since our WLM is monitoring a large number of home networks, active DNS queries are done once for every IoT type (and not for every IoT device), thus reducing the load on the DNS server. 

\subsubsection{Analyzing traffic of unidentified devices}
\label{sec:controlplane_analyze}
Recall that for devices without MUD files, the control plane receives the list of  domains and endpoints the device communicates with. The control plane compares (in an efficient manner) this list of to all known whitelists/MUD files. If the device matches a whitelist of some IoT device type $X$, we associate these marked packets with $X$ by rewriting the corresponding rule at Table 2 to point to Table $X$ (see Figure~\ref{table:tables}) and by that ensuring that the device  continues to comply with $WL_X$. 

Otherwise, the device is either an IoT device that has never been seen before (and therefore a new whitelist is created for it) or is not an IoT device. \anat{ A copy of the traffic is sent to a control application, that identifies if the device is IoT or not IoT (using techniques such as \david{in} \cite{Technical_report}). If the device is a new IoT device with no MUD \david{file}, we send \david{the traffic (or part of the traffic)} to \david{a} control application that algorithmically learns the whitelist corresponding to the device. As mentioned before, the system described in this paper provides only the infrastructure for MUD file generation and the exact algorithmic way to do so (e.g., by variation of~\cite{DBLP:journals/corr/abs-1804-04358}) is out of  scope. \david{Finally, after a MUD file is provided, we} 
install the corresponding whitelist on a separate Table $X'$, and associate the device with $X'$. If the device is not an IoT device, the control plane instructs the CPE to remove the DSCP mark from the device's packets, and therefore, all subsequent packets will arrive at the VNF unmarked (see Section~\ref{sec:control_acs}).
}

\ignorenoms{
(we omit the details of how we distinguish between these two cases due to space limitation). In the former case, we algorithmically learn the whitelist corresponding to the device (e.g., by using the algorithm described in~\cite{DBLP:journals/corr/abs-1804-04358}), install it on a separate Table $X'$, and associate the device with $X'$. In the latter case, the control plane instructs the CPE to remove the DSCP mark from the device's packets, and therefore, all subsequent packets will arrive at the VNF unmarked (see Section~\ref{sec:control_acs}).}

\subsubsection{Remotely Configuring CPEs}
\label{sec:control_acs}

One of the most important parts of the control plane is the communication between the VNF and the CPE,  to configure the latter. Such a configuration can be done in various ways, depending on the CPE capabilities. However, in the vast majority of cases, the CPE supports the TR-069 protocol for its remote configuration and ISPs use this protocol to configure the CPE connected to their network. Thus, our description will focus on this protocol. Similar operations (and configurations) can be done using SNMP and proprietary management protocols. We note that once configuring a CPE, 
this configuration remains persistent (unless it is factory reset).

TR-069~\cite{tr69} is a client-server protocol, where the CPE acts as a client and periodically initiates connections to an \emph{Auto-Configuration Server} (ACS). When a connection is established, the ACS sends configuration requests to the CPE, which are executed one by one. The ACS also exposes a north-bound interface, so that management tools and applications can communicate with it. Thus, our VNF, like any other TR-69--based application, communicates with the ACS (in our case, the ISP's ACS), which in turn, passes the requests to the CPE. An important feature of TR-069 is the ability to set a specific parameter for ``active notification''. In such a case, if the parameter  has been modified by an external cause (a cause other than the ACS
itself), the CPE initiates a connection immediately to the ACS and notifies it on the parameter change. We will use this feature later to detect new devices that join the customer's network. 
The TR-069 supports data models that represent the entities that need to be configured. CPEs are modeled using the Internet Gateway Device (IGD) Data Model (TR-98)~\cite{tr98} and Device Data Model (TR-181)~\cite{tr181}, and our VNF  uses only mandatory parameters within these models. 

Our VNF initiates configuration requests upon the following events:

\smallskip
\noindent
{\bf New customer is added to the network:} The external IP address of the customer is extracted (so that the VNF can identify connections from that customer). In addition, the respective TR-98 parameter is set to ``active notification'', so that future changes to the customer's external IP address will be reported automatically. Moreover, the VNF extracts basic details on all hosts that are connected to the CPE, and for each host, it adds a mapping between its MAC address and a DSCP mark. 

Also, it adds a default DSCP mark (see Section~\ref{sec:dp_guard}) and sets active notifications for the number of connected hosts (to detect hosts that join the customer's network). 

Finally, in many settings,  devices that are connected to the network are configured (through DHCP) to use the CPE as their DNS resolver. This implies that each DNS request is terminating in the CPE, which, in turn, issues another DNS request to ISP's DNS resolver, if needed. This implies that DNS requests arrive at the CPE unmarked. Since our data-plane tracks DNS queries, and since it is important to associate each query with the device that issued it, we change this configuration, so that each device communicates directly with the ISP's DNS resolver. 

\smallskip
\noindent
{\bf An external IP address is changed:} Recall that this parameter is marked for active notification so that as soon as this happens, the CPE notifies the ACS (and then our VNF) on this change. The VNF then updates its internal state to monitor connections/packets from the new external IP address rather than the old one (namely, in Table 1 of the pipeline described in Figure~\ref{table:tables}). 

\noindent
{\bf New device joins the network:} As the number of hosts connected to the CPE is marked for active notification, the VNF is notified immediately upon such an event. Then, basic details of the device (such as its MAC address, hostname, the medium of connection, etc.) are extracted from the CPE, and the CPE is configured to mark all packets transmitted from this device. The VNF is configured to monitor these packets as well as to associate all packets with default mark to this device (to compensate on the time until the DSCP mark comes into effect). 

As MUD files are specified by IoT devices through DHCP, we check the specific DHCP option (available in the Device data model) for the URL of the MUD file. If such exists, we check if the MUD file is already installed in the VNF and start  monitoring the device by associating it to the correct table (see Table 2 in Figure~\ref{table:tables}). If the MUD file is not installed, we download, parse, and install it as a new table in the VNF. 
When no MUD file is specified, all packets transmitted from the new device are marked packets of an unidentified device.
We note that we consider a device as new only on the first time it connects to the CPE. If the device disconnects from the network and re-connects again, the CPE still lists it under its hosts (as inactive hosts), implying DSCP marks are already available, and no further action is required. 

\noindent
{\bf A device is classified as a non-IoT device by the VNF:} In this case, the DSCP mark on packets transmitted for this specific device should be removed, so that the VNF will stop monitoring the device by discarding the unmarked packets. This is done by setting the \texttt{DSCPMark} within the respective classification object to value -1, indicating that the CPE should not change the DSCP value of packets.

%% file: P2P.tex
\section{Extending MUD for P2P communication}
 \label{sec:p2p}
 
 One important limitation of MUD files is that it does not cover cases where the IoT device is a server and its users act as clients, whose IP addresses can change frequently. Such P2P communication is common when the IoT device and its users are in the same LAN (namely, the same customer network) or when a low latency connection between the device and its users is required (e.g., for improving QoS of a video streaming): 
\david{
In our \anatv{study, of IoT traffic logs\cite{Sivanathan2017}, \cite{cicids2017} , \cite{cicids2012} and IoT devices in our lab, out of 42 IoT devices}, five IoT devices\footnote{The  devices are Belkin Wemo switch, Belkin Wemo Motion Sensor, 
Motorola Hubble,
TP-Link Day Night Cloud camera, and
Samsung Smart Cam. All these cameras use STUN~\cite{rosenberg2003stun}  to allow incoming connections to the home networks, as the cameras are typically behind NAT. } and one IoT alarm system act as servers. We note that IoT P2P communication is  less frequent nowadays, as most IoT devices communicate with their users through a proxy in the cloud (and are covered by our NFV-based solution as described in Section~\ref{sec:sys}). Yet, recent devastating IoT attacks (such as Mirai~\cite{MIRAI}) targeted exactly such devices.
}
 
In most cases, users access IoT devices through a dedicated application, and therefore, we will use the term \emph{IoT app} to refer to these users and distinguish them from users of our security system. Furthermore, the device on which the IoT app is installed is called \emph{{IoT app device (IAD)}}. 
 
 The main problem of MUD for P2P communication is that the IoT app that is allowed to access a certain IoT device, is typically not allowed to access all IoT devices of the same type (e.g., user of a camera of type $X$ should not access other cameras of type $X$ that it does not own). Thus, without additional mechanisms, the device manufacturer cannot specify in the MUD file any details on the IoT app. One common approach to cope with that, similar to what was shown in \cite{Hamza:2018:CMP:3229565.3229571}, is to allow incoming traffic from any IP in the internet (i.e., "*"); naturally, such a whitelist has close-to-zero precision.  
 
 We have taken a different approach and extend the MUD specification to allow a \emph{secondary virtual manufacturer (SVM)}, whose sole purpose is to deal with P2P communication. Notice that this virtual manufacturer may deal with all P2P communication in the network (regardless of the manufacturer of IoT device). Adding such a secondary virtual manufacturer requires a minimal change in the specification and minimal cooperation from the primary (physical) manufacturer: In the MUD file, which the primary manufacturer provides to the WLM  (often called MUD manager in this context), it should specify that its device supports direct connection. This can be done by inserting a unique placeholder (in our implementation, \texttt{\$owner-unique-domain\$}) to the file, e.g., by adding the following rule:
\vspace{0.5em} 

 \begin{minipage}[h!]{\textwidth}
 \begin{verbatim}
  "ipv4": {
   	"ietf-acldns:src-dnsname": 
   	      "$owner-unique-domain$",
    "protocol": 6
  }
 \end{verbatim}
\end{minipage}
Upon encountering the unique placeholder, the WLM permanently replaces it with the specific domain (or domains) provided by the SVM, as will be described next. Upon such replacement, the WLM is left with a regular MUD file and can continue its normal operations without any change. 

\subsection{Deploying WLM and WLE for P2P Communication} 
As P2P communications include local traffic and, in any case, the MUD file of every device (and not every device type) is different, the WLM and WLE should be deployed within the Customer Premise Equipment (CPE); namely, the home gateway router. This can be done by installing a software on the CPE device, which can run either as a process or as a container in this machine.  CPE-based solution implements WLM and WLE on the same box, but can be  deployed as different processes, where WLE can sometimes use available hardware to enhance its performance.

CPE-based deployments have the best visibility as it can observe and block unauthorized internal traffic between devices in the same network. Moreover, as typically there is no NAT between the CPE and the IoT devices, it is easy to associate packets to their corresponding IoT device (no need for DSCP marking, etc.). On the other hand, WLM must be able to resolve all domain names in all its MUD files to maintain the correct mapping to IP addresses. This can be done either by extracting the addresses from DNS responses (which, in turn, requires deep packet inspection of packets and incurs additional overhead) or periodically performing all the required DNS queries.  

\begin{figure}
\includegraphics[clip, trim=0.1cm 0.2cm 0.1cm 1.6cm, width=0.49\textwidth]{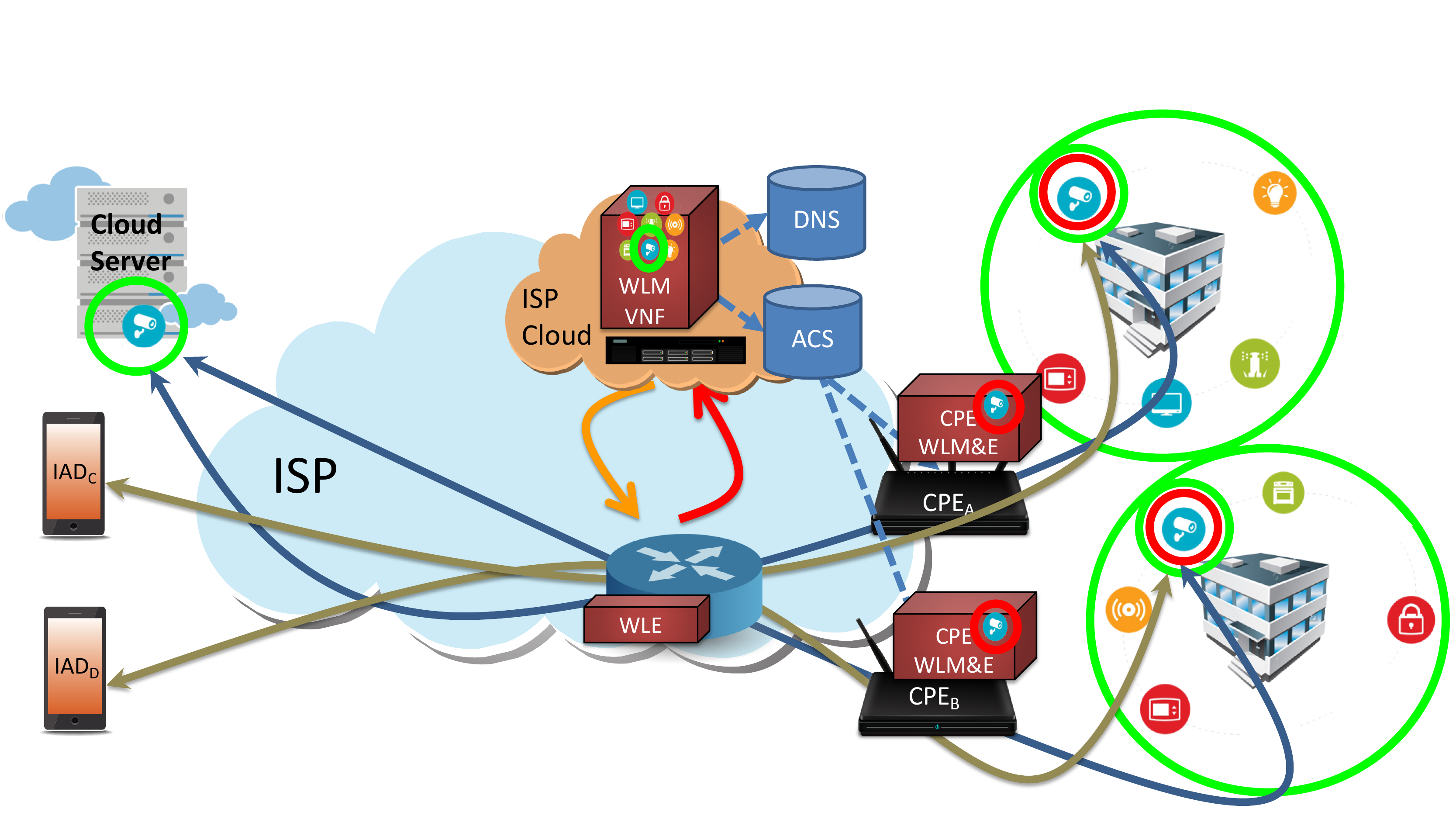}
\caption{\label{fig:framework_cpe}  Illustration of our framework, where WLM\&E component in the CPE is responsible on P2P and local communication, while the VNF-based WLM is responsible for all other communication.  In this illustration, the same camera is installed in two homes, where $\mbox{IAD}_C$ ($\mbox{IAD}_D$) connects directly to the camera at  $\mbox{CPE}_A$ ($\mbox{CPE}_B$). A whitelist with the IP address of $\mbox{IAD}_C$ ($\mbox{IAD}_D$) is installed in  $\mbox{CPE}_A$ ($\mbox{CPE}_B$). A single whitelist in the VNF-based WLM monitors all other communications for both cameras. }
\end{figure}

Recall that we envision a hybrid deployment, which has CPE-based component for P2P and internal communication of IoT devices and NFV-based component for all other traffic, thus achieving the best of both worlds. \david{ See Fig.~\ref{fig:framework_cpe}.
For IoT devices that have both types of communication, the whitelist is spread across the two components, which might cause conflicts \anatv{in a naive implementation}: connections that are approved by the whitelist in the CPE are later examined by the NFV-based WLM and are blocked. To avoid such conflict, we make sure that the CPE-based WLE component reset the DSCP value of all packets matching its whitelist (implying they will not be inspected or blocked by 
the NFV-based components).\footnote{This is done in two steps: the WLE at the CPE marks the packets by a designated DSCP marking, and the CPE, as instructed by the NFV-based WLM through TR-69, reset the DSCP marks of such packets regardless of their source MAC address.} 
}

\subsection{The Secondary Virtual Manufacturer (SVM) } 

The SVM is responsible for generating specific domain names for IADs and keeping a correct mapping between these domain names and their corresponding IP addresses.  The main challenge is that, typically, IADs' IP addresses are changing very frequently (e.g., when the IAD is a mobile device and it moves between networks) implying the SVM should \emph{track} the relevant IADs and update the corresponding mapping. In a nutshell, this is done by installing an \emph{SVM tracking application} on the IAD, which  consistently reports the IP address of the IAD (and other metadata). This data is sent to the \emph{SVM's mapping service} (deployed in the cloud) which updates the DNS record that maps between the IAD's IP address to a unique domain. We note that, as we work with DNS, changes in IP addresses do not cause any change in the normal operations of the WLM, which include domain name resolutions and coping with DNS changes. 

Specifically, in our implementation, the SVM owns a parent domain and given each IAD a unique randomly-generated sub-domain under this domain. The SVM maintains  the authoritative DNS server for the parent domain (and hence, all its sub-domains) to ensure that changes to DNS records, generated by the \emph{SVM's mapping service}, are reflected in subsequent DNS queries. Yet, one major issue that can be caused due to this system architecture outdated  resolution due to stale cached records in one of the DNS revolvers. This problem can be circumvented in standard techniques that include adding random value as a sub-domain so that each DNS query will bypass all caching mechanisms or pointing the WLM to the SVM's \anatv{authoritative} DNS  server. \anatv{ The authoritative DNS server will then remove the random value and return update answer.  }

\begin{figure}[t!]
\includegraphics[clip, trim=1.4cm 1.3cm 1.5cm 1.25cm, width=0.5\textwidth]{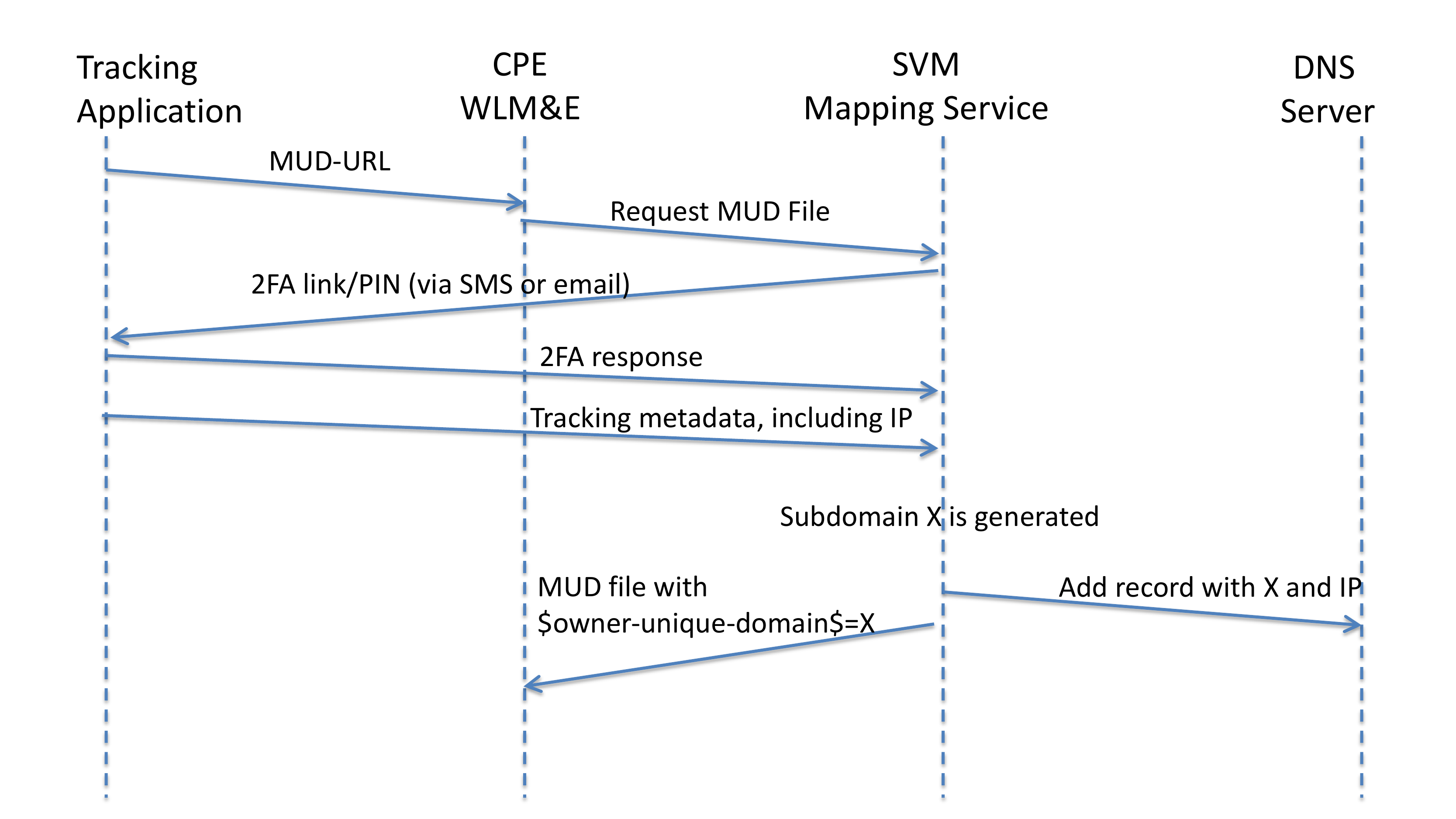}
\caption{\label{fig:p2psignup}  Illustration of the sign-up process of a IAD with the SVM and the corresponding CPE. The sign-up process must be completed when the IAD is connected through the CPE.  \vspace{-0.5cm}}
\end{figure}

The most involved operation of the SVM is when a new user sign-up with the service, which is Illustrated in Fig.~\ref{fig:p2psignup}.
A user who wants to use this protection needs to install the SVM tracking application and create an account with the SVM first. To enable two-factor authentication, users are required to provide either an email address and/or phone number. During the sign-up process a new unique sub-domain and identifier to the account are generated. To correlate the newly-correlated account with a specific LAN, we require that the SVM tracking application will be connected, in the first time, during that LAN. The operations continue in the following steps, where the SVM tracking application acts as a (virtual) IoT device, while the SVM mapping service acts as its manufacturer:
(i) The SVM tracking application broadcasts a MUD URL, which includes the SVM parent domain and the identifier of the newly created account; 
(ii)	The WLM fetches this MUD URL by contacting its manufacturer as specified in the URL. This request goes to the SVM's mapping service; 
(iii)	The SVM's mapping service extracts the account identifier and initiates a two-factor authentication with the user, using the email address and/or phone number provided when the account was created; 
(iv)	Upon successful completion of the two-factor authentication, the mapping service replies to WLM with the MUD file that includes the account unique domain (namely, the value of \texttt{\$owner-unique-domain\$});   
(v)	The WLM extracts this unique domain from the MUD file, replaces all pending MUD files with \texttt{\$owner-unique-domain\$}, and save the domain name for later usage.

\begin{figure}[t!]
\includegraphics[clip, trim=1.4cm 1.3cm 1.5cm 1.25cm, width=0.5\textwidth]{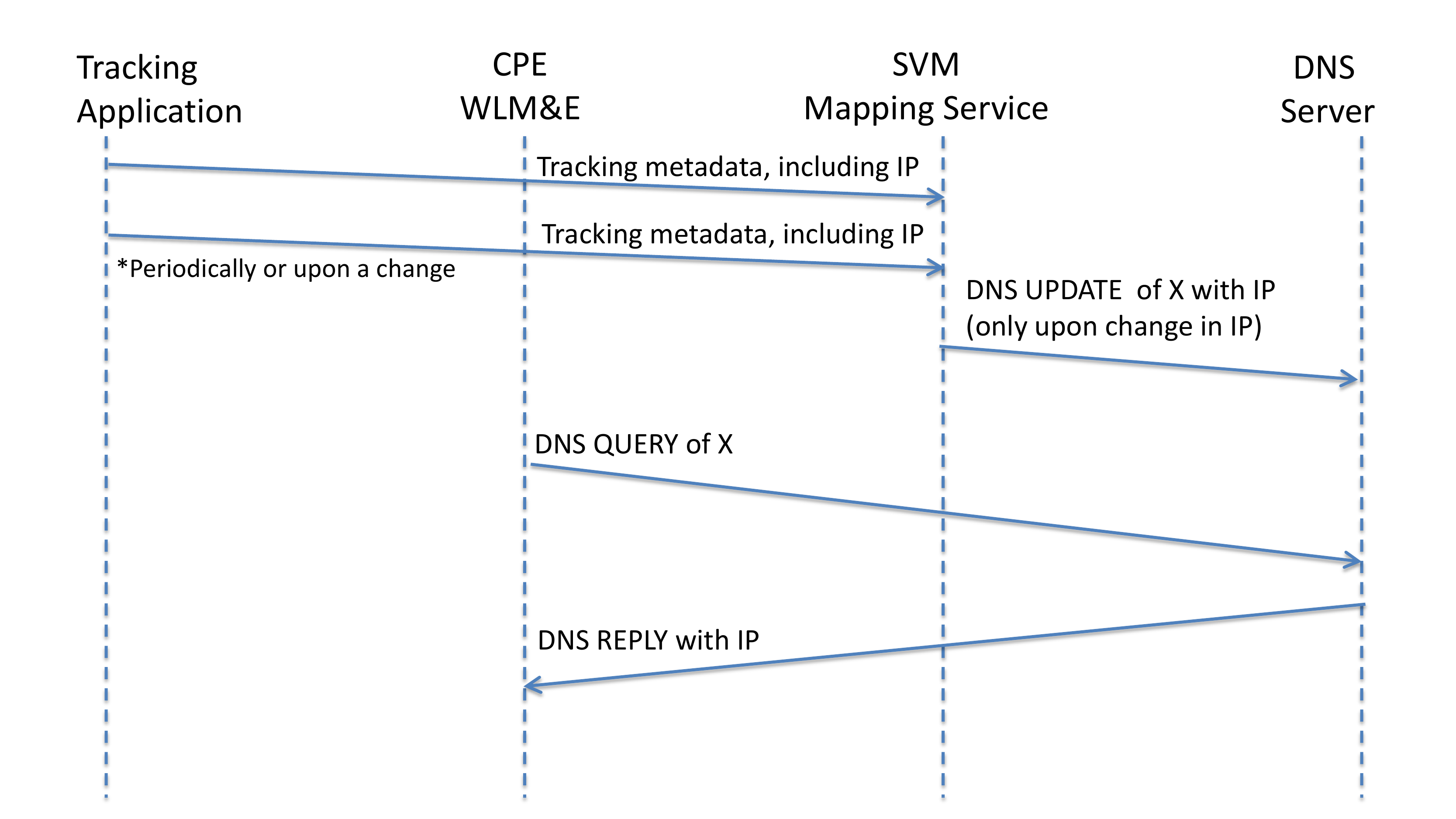}
\caption{\label{fig:p2pregular}  Illustration of the messages sent after the initial sign-up process. Notice that the IoT device itself is oblivious to the system and continues its operation without any change. }
\end{figure}

We note that after an account is created, the SVM tracking application reports to the SVM mapping service whenever the external IP address of the IAD is changed (or every 15 minutes, as a keep-alive message). The message sent includes the account identifier, the unique domain, the external and internal IP addresses. If there is a change, the mapping service simply updates the corresponding DNS record in the authoritative DNS server. See a description of  these operations in Fig.~\ref{fig:p2pregular}.

\subsection{Dealing with Internal traffic}
When the IoT app is in the same LAN as the IoT device, mapping to IP addresses  is not required, and instead one can use other information available within the LAN to define whitelist rules in the MUD file (e.g., the IAD's MAC address). This can be generated in exactly the same process as described above, with an additional place-holder.  Note however that in some cases, mostly in the presence of extenders within the LAN that rewrite MAC addresses, the IAD's internal IP address is needed; in this case, we configure the tracking application to send both internal and external addresses and configure two DNS records for that IAD. We note the case where communication between different IoT devices in the same LAN should be monitored was recently solved in~\cite{ranganathan2019soft}.  Reducing the surface of the attack also inside the LAN is extremely important, since there are many attacks that abuse local devices to internally attack IoT devices \cite{acar2018web}, or to leak data from them~\cite{fishtank}.

%% file: implement.tex
\label{sec:implement}

We have implemented our system, and deployed a proof-of-concept in a large national-level ISP network.

\ignore{\input{dashboard}
}

Specifically, the data-plane was implemented using Open vSwitch (OVS) version 2.8.1~\cite{ovs} with OpenFlow 1.3. The control plane was implemented as applications (in Python) over Ryu---a common open-source OpenFlow controller~\cite{ryu}. 
Our implementation leverages on both the caching capabilities of OVS, its supports of DPDK, and the pipelined architecture of OVS and OpenFlow, where packets traverse multiple tables, each contains several rules before they are classified.

DNS traffic and traffic from unidentified devices (namely, devices with no corresponding MUD file) is sent to the controller, on top of which there is an application that tries to decide whether an unidentified device is an IoT device or not, to match unidentified device activity with an existing MUD file, to derive new whitelists, and to continue monitor changes in whitelists. 
Identified devices' traffic that is destined for IP addresses not within the whitelist/MUD file is forwarded to another application, which re-verifies that the traffic is indeed legitimate and issues an alert to the customer and an ACL to ISP, which can block this specific connection. The latter application communicates with the applications on top of the controller (e.g., to force active DNS queries) using RESTful APIs. We have also used the built-in OVS counters to track the activity on each IoT device and present it to the customers and the ISP.    

We note that while we have implemented and tested our solution with OVS, as the solution is OpenFlow based and does not use any proprietary fields of OVS, one can easily migrate it to use any (physical or virtual) switch that supports OpenFlow 1.3 and above. 

In our PoC deployment, we do not have access to the ISP's ACS servers. Thus, we have deployed our own ACS server, running open-source GenieACS 1.1~\cite{genieacs}, and configured the CPEs to communicate with it. As GenieACS does not support active notification setting yet, we have shortened the \texttt{inform} interval and manually refresh the corresponding TR-98 fields, to capture the monitored events (mainly, new device joins the customer's network for the first time, or a new port is opened in the CPE). Our VNF communicates with GenieACS through GenieACS's north-bound interface. 

We have four VTech NB-403 CPEs, running for several months in different locations within the ISP network. The ISP routes all traffic to/from these CPEs through our VNF, which runs as a virtual machine on top of one of the ISP's VMware ESXi hypervisors. 

As for the extension to deal with P2P communication, we have implemented the tracking application and installed it with dozens of volunteers and checked some properties of the solution. We highlight here some observation from this experiment: First, DNS records were updates 15 times a day on average (with a maximum of 40 times).  DNS record changes happen when IAD is switching between a WiFi and cellular connections or between different WiFi networks. On the other hand, the same IP address might belong to many records (namely, when IADs using the same WiFi networks or the same cellular provider, in case that provider use Carrier-Grade NAT~\cite{8486223}); this implies that users within the same network of the IAD might bypass our security mechanism (or any other MUD-based solution). Finally, in some cases DNS records must contain more than one IP address, since in some enterprise networks, due to load-balancing of  outgoing connections, every IAD connection may receive different (external) IP address. 

%% file: Discussion.tex
\label{sec:discussion}
In this paper, we have presented a system to provide IoT security through a VNF deployed within the ISP. Our system is built upon the MUD specification, in which each device has a whitelist that describes its legitimate communication patterns, and any deviation from this whitelist is blocked. Our system may use a MUD file provided by the device (through DHCP, as specified in the MUD RFC). It also provides the infrastructure to derive MUD files from unidentified devices' traffic   (e.g., by running the algorithm described in~\cite{DBLP:journals/corr/abs-1804-04358}) and to identify devices (that do not provide their MUD files by themselves) by matching their traffic to  existing MUD files (or whitelists).

Furthermore, as our system is deployed in a centralized location, monitoring a large number of home networks simultaneously, it is in an excellent position to detect global phenomena such as outgoing and incoming DDoS attacks. Our future work includes designing and deploying such mechanisms within our system.